\documentstyle[11pt,
epsfig]{article}

\begin{document}

\relax
\renewcommand{\theequation}{\arabic{section}.\arabic{equation}}

\def\be{\begin{equation}}
\def\ee{\end{equation}}
\def\bs{\begin{subequations}}
\def\es{\end{subequations}}
\def\calm{{\cal M}}
\def\calk{{\cal K}}
\def\lx{\lambda}
\def\sx{\sigma}
\def\ex{\epsilon}
\def\Lx{\Lambda}

\newcommand{\phit}{\tilde{\phi}}
\newcommand{\mt}{\tilde{m}}
\newcommand{\nt}{\tilde{n}}
\newcommand{\Nt}{\tilde{N}}
\newcommand{\Bt}{\tilde{B}}
\newcommand{\Rt}{\tilde{R}}
\newcommand{\rt}{\tilde{r}}
\newcommand{\mut}{\tilde{\mu}}
\newcommand{\mub}{\bar{\mu}}
\newcommand{\vrm}{{\rm v}}
\newcommand{\tl}{\tilde t}
\newcommand{\ttt}{\tilde T}
\newcommand{\rhot}{\tilde \rho}
\newcommand{\ptt}{\tilde p}
\newcommand{\drho}{\delta \rho}
\newcommand{\drhot}{\delta {\tilde \rho}}
\newcommand{\dchi}{\delta \chi}
\newcommand{\A}{A}
\newcommand{\B}{B}
\newcommand{\mmu}{\mu}
\newcommand{\mnu}{\nu}
\newcommand{\ii}{i}
\newcommand{\jj}{j}
\newcommand{\jl}{[}
\newcommand{\jr}{]}
\newcommand{\ml}{\sharp}
\newcommand{\mr}{\sharp}

\newcommand{\da}{\dot{a}}
\newcommand{\db}{\dot{b}}
\newcommand{\dn}{\dot{n}}
\newcommand{\dda}{\ddot{a}}
\newcommand{\ddb}{\ddot{b}}
\newcommand{\ddn}{\ddot{n}}
\newcommand{\pa}{a^{\prime}}
\newcommand{\pn}{n^{\prime}}
\newcommand{\ppa}{a^{\prime \prime}}
\newcommand{\ppb}{b^{\prime \prime}}
\newcommand{\ppn}{n^{\prime \prime}}
\newcommand{\fda}{\frac{\da}{a}}
\newcommand{\fdb}{\frac{\db}{b}}
\newcommand{\fdn}{\frac{\dn}{n}}
\newcommand{\fdda}{\frac{\dda}{a}}
\newcommand{\fddb}{\frac{\ddb}{b}}
\newcommand{\fddn}{\frac{\ddn}{n}}
\newcommand{\fpa}{\frac{\pa}{a}}
\newcommand{\fpb}{\frac{\pb}{b}}
\newcommand{\fpn}{\frac{\pn}{n}}
\newcommand{\fppa}{\frac{\ppa}{a}}
\newcommand{\fppb}{\frac{\ppb}{b}}
\newcommand{\fppn}{\frac{\ppn}{n}}

\newcommand{\dphi}{\delta \phi}
\newcommand{\at}{\tilde{\alpha}}
\newcommand{\pt}{\tilde{p}}
\newcommand{\Ut}{\tilde{U}}
\newcommand{\rhb}{\bar{\rho}}
\newcommand{\pb}{\bar{p}}
\newcommand{\pbb}{\bar{\rm p}}
\newcommand{\kt}{\tilde{k}}
\newcommand{\kb}{\bar{k}}
\newcommand{\wt}{\tilde{w}}

\newcommand{\dA}{\dot{A_0}}
\newcommand{\dB}{\dot{B_0}}
\newcommand{\fdA}{\frac{\dA}{A_0}}
\newcommand{\fdB}{\frac{\dB}{B_0}}

\def\be{\begin{equation}}
\def\ee{\end{equation}}
\def\bs{\begin{subequations}}
\def\es{\end{subequations}}
\newcommand{\een}{\end{subequations}}
\newcommand{\ben}{\begin{subequations}}
\newcommand{\beq}{\begin{eqalignno}}
\newcommand{\eeq}{\end{eqalignno}}

\def \lta {\mathrel{\vcenter
     {\hbox{$<$}\nointerlineskip\hbox{$\sim$}}}}
\def \gta {\mathrel{\vcenter
     {\hbox{$>$}\nointerlineskip\hbox{$\sim$}}}}

\def\g{\gamma}
\def\mpl{M_{\rm Pl}}
\def\ms{M_{\rm s}}
\def\ls{l_{\rm s}}
\def\l{\lambda}
\def\m{\mu}
\def\n{\nu}
\def\a{\alpha}
\def\b{\beta}
\def\gs{g_{\rm s}}
\def\d{\partial}
\def\co{{\cal O}}
\def\sp{\;\;\;,\;\;\;}
\def\r{\rho}
\def\dr{\dot r}

\def\e{\epsilon}
\newcommand{\NPB}[3]{\emph{ Nucl.~Phys.} \textbf{B#1} (#2) #3}   
\newcommand{\PLB}[3]{\emph{ Phys.~Lett.} \textbf{B#1} (#2) #3}   
\newcommand{\ttbs}{\char'134}        
\newcommand\fverb{\setbox\pippobox=\hbox\bgroup\verb}
\newcommand\fverbdo{\egroup\medskip\noindent%
                        \fbox{\unhbox\pippobox}\ }
\newcommand\fverbit{\egroup\item[\fbox{\unhbox\pippobox}]}
\newbox\pippobox
\def\tr{\tilde\rho}
\def\lb{w}
\def\bbox{\nabla^2}
\def\mt{{\tilde m}}
\def\rct{{\tilde r}_c}

\def \lta {\mathrel{\vcenter
     {\hbox{$<$}\nointerlineskip\hbox{$\sim$}}}}
\def \gta {\mathrel{\vcenter
     {\hbox{$>$}\nointerlineskip\hbox{$\sim$}}}}

\noindent
\begin{flushright}

\end{flushright} 
\vspace{1cm}
\begin{center}
{ \Large \bf Static Configurations of 
Dark Energy and Dark Matter\\}
\vspace{0.5cm}
{N. Brouzakis and N. Tetradis} 
\\
\vspace{0.5cm}
{\it University of Athens, Department of Physics, \\
University Campus, Zographou 157 84, Athens, Greece.} 
\\
\vspace{1cm}
\abstract{
We study static configurations of dark matter coupled to a scalar field
responsible for the dark energy of the Universe. The dark matter is modelled
as a Fermi gas within the Thomas-Fermi approximation. The mass of the
dark matter particles is a function of the scalar field.
We analyze the profile of the dark matter halos in galaxies. 
In this case our framework is equivalent to the model of the isothermal sphere.
In the presence of 
a scalar field, the velocity of a massive object orbiting 
the galaxy is not of the order of the typical velocity of
the dark matter particles, as in the conventional picture. Instead, it is
reduced by a factor that quantifies the 
dependence of the dark matter mass on the scalar field.
This has implications for dark matter searches.
We derive new solutions of the Einstein equations which describe
compact objects composed of dark matter. 
Depending on the scale of the dark matter mass,
the size of these objects can vary between microscopic scales and 
cosmological distances. 
We determine the 
mass to radius relation and discuss the similarities with
conventional neutron stars and exotic astrophysical objects.
\\
\vspace{1cm}
} 
\end{center}

\newpage

\section{Introduction}
\label{intro}
\setcounter{equation}{0}

Our present understanding of the history of the Universe
assumes the presence of various contributions to its energy content.
The total energy density is believed to take the critical value that
results in a spatially flat Universe. 
The most accessible energy component is baryonic matter, which accounts for
$\sim 5\%$ of the total energy density. A component that has
not been directly observed is dark matter: a pressureless fluid that is
responsible for the growth of cosmological pertrurbations through
gravitational instability. Its contribution to the total energy density is
estimated at 
$\sim 25\%$. The dark matter is expected to become more numerous 
in extensive halos, that
stretch up to 100--200 kpc from the center of galaxies \cite{peeblesb}.
%Ongoing experiments are trying to detect the 
%dark matter particles in the halo of the Milky Way \cite{cline}. 
The component with the biggest contribution to the energy density has
an equation of state similar to that of a cosmological constant. 
The ratio $w=p/\rho$ is negative and close to $-1$. 
(For a review, see ref. \cite{peeblesrev}.)
This component is responsible for $\sim 70\%$ of the total energy density 
and induces
the observed acceleration of the Universe \cite{supernova}. 

The difficulty with explaining the very small value of the 
cosmological
constant that could induce the present acceleration has
motivated the suggestion that this energy component is time dependent
\cite{wetdil,peeblesold}. In the simplest realization, it is connected to
a scalar field $\phi$ with a very flat potential. The vacuum energy 
associated with this field is characterized as dark energy and drives
the acceleration. 
If such a field affects
the cosmological evolution today, its effective mass must be of the order 
of the Hubble scale, or smaller. 

It is conceivable 
that there is a coupling between dark matter and
the field responsible for the dark energy \cite{wetcosmon}. 
In such a scenario it may be
possible to resolve the coincidence problem, i.e. the reason behind
the comparable present 
contributions from the dark matter and the dark energy to the
total energy density. 
The construction of specific models 
is still in the early stages \cite{amendola}--\cite{spec}. 
Phenomenological difficulties arise because of 
the required flatness of the potential of the scalar field.
The coupling between
the field and the dark matter could lift the
flatness of the potential through radiative corrections. 
We assume that at the level of the effective potential the 
required flatness is maintained, possibly guaranteed by some symmetry 
\cite{radiative}.

The presence of an interaction between dark matter 
and the scalar field responsible for the dark energy
has consequences that are potentially observable.
The cosmological implications depend on the form of the coupling,
as well as on the potential of the field \cite{amendola}. 
If the scale for the field mass is
set by the present value of the Hubble parameter, then
the field is effectively massless at distances of 
the order of the galactic scale. 
Its coupling to the dark matter particles results in a long range 
force that can affect the details of structure formation 
\cite{largescale,maccio,peeblessim}.

The attraction between
dark matter particles mediated by the scalar field
may lead to the formation of 
dense compact objects composed primarily of dark matter.
We are interested in static solutions of the Einstein
equations that describe such objects.
Pure scalar field configurations, coupled to gravity,
are known. An explicit analytical solution
has been derived in ref. \cite{wetl}.
This solution, however, has
a naked singularity at the origin. 
In the present work we show
that the addition of fermions can eliminate the singularity and
result in a configuration that has a natural physical interpretation.

In the following section we develop the necessary
formalism for our study.
We assume that the
interaction between dark matter and dark energy takes the
form of an effective field-dependent 
mass term for the dark matter particles.
We derive the Einstein equations and the equation of motion of 
the scalar field.
In order to derive an equation of state for the dark matter, we 
model it as a Fermi gas in the Thomas-Fermi
approximation. 
In section \ref{known} we summarize known solutions of the system
of equations of section \ref{formalism}. We also discuss a
class of solutions that gives an
approximate description of galaxy halos in the presence of the scalar field.
In section \ref{astro} we derive new solutions that describe compact objects 
composed of dark matter and held together by
the scalar interaction. Section \ref{conclusions} 
contains a summary and our conclusions.

\section{Formalism}
\label{formalism}
\setcounter{equation}{0}

\subsection{The basic expressions}

We assume that the dark matter consists of a gas of weakly interacting 
particles. The mass $m$ of the particles depends on
the value of a slowly varying classical scalar field $\phi$ \cite{wetcosmon}. 
For classical particles, the action of the system can be written as (see ref. 
\cite{peeblesfar}
and references therein)
\begin{equation}
{\cal S}=\int d^4x \sqrt{-g}
\left(M^2 R-\frac{1}{2}
\frac{\partial \phi}{\partial x^\mu}
\frac{\partial \phi}{\partial x^\nu}
g^{\mu\nu}-U(\phi) \right)
-\sum_i \int m(\phi(x_i))d\tau_i,
\label{action} \end{equation}
with $d\tau_i=\sqrt{
-g_{\mu\nu}(x_i)dx^\mu_idx^\nu_i}$ and
the second integral taken over particle trajectories. 
Variation of the action with respect to $\phi$ results in the
equation of motion \cite{peeblesfar,damour}
\be
\frac{1}{\sqrt{-g}}\frac{\partial}{\partial x^\mu}
\left(\sqrt{-g}\,\,g^{\mu\nu}\frac{\partial \phi}{\partial x^\nu}
\right)=
\frac{dU}{d\phi}+W,
\label{eomphi} \ee
where
\be
W=\frac{1}{\sqrt{-g}}\sum_i\int d\tau_i \,\, 
\frac{dm(\phi(x_i))}{d\phi} \, \delta^{(4)}(x-x_i)
=-\frac{d\ln m(\phi(x))}{d\phi}\,\, T^\mu_{~\mu}.
\label{press} \ee
The energy-momentum tensor associated with the gas of particles
is 
\be
T^{\mu\nu}=\frac{1}{\sqrt{-g}} 
\sum_i \int d\tau_i \,\, m(\phi(x_i))\,\, 
\frac{dx_i^\mu}{d \tau_i}\frac{dx_i^\nu}{d \tau_i}
\delta^{(4)}(x-x_i).
\label{emtg} \ee
In the following we shall look for solutions of 
eqs. (\ref{eomphi}), (\ref{press}) 
employing an approximation for the form of the energy-momentum
tensor $T^{\mu\nu}$. We shall assume that it takes the 
diagonal form $T^\mu_{~\nu}={\rm diag} (-\rho,p,p,p)$.

We are interested in stationary, spherically symmetric configurations
of the system. 
We consider a metric of the form
\be
ds^2=-B(r)dt^2+r^2(d\theta^2+\sin^2\theta\, d\phi^2)+A(r)dr^2.
\label{metric} \ee
The Einstein equations are
\begin{eqnarray}
\frac{1}{r^2}\frac{1}{A}-\frac{1}{r^2}-\frac{1}{r}\frac{A'}{A^2}
&=&\frac{1}{2M^2}\left( -\frac{1}{2A}\phi'^2-U(\phi)-\rho\right)
\label{ein1} \\
-\frac{1}{2r}\frac{A'}{A^2}+\frac{1}{2r}\frac{B'}{AB}
-\frac{A'B'}{4A^2B}-\frac{B'^2}{4AB^2}+\frac{B''}{2AB}
&=&\frac{1}{2M^2}\left( -\frac{1}{2A}\phi'^2-U(\phi)+p\right)
\label{ein2} \\
\frac{1}{r^2}\frac{1}{A}-\frac{1}{r^2}+\frac{1}{r}\frac{B'}{BA}
&=&\frac{1}{2M^2}\left( \frac{1}{2A}\phi'^2-U(\phi)+p\right),
\label{ein3} 
\end{eqnarray}
where a prime denotes a derivative with respect to $r$.
The scale $M$ is defined as $M=(16\pi G)^{-1/2}$, where $G$ is Newton's 
constant.
The equation of motion (\ref{eomphi}) for the field $\phi$ becomes
\be
\phi''+\left(\frac{2}{r}-\frac{A'}{2A} +\frac{B'}{2B} \right) \phi'=
A\left[ \frac{dU}{d\phi}+\frac{d\ln m(\phi)}{d \phi}\,\,(\rho-3p)\right]. 
%= A\left[ \frac{dU}{d\phi}-\frac{\partial p}{\partial \phi}\right].
\label{eomphii} \ee
Combining eqs. (\ref{ein1}), (\ref{ein3}), (\ref{eomphii}), 
we can eliminate $A'$ and $B'$ from the equation of motion of the field $\phi$
\be
\phi''+\left[\frac{1+A}{r}-\frac{1}{2M^2}Ar\left( U(\phi)
+\frac{1}{2}(\rho-p) 
\right) \right] \phi'=
A\left[ \frac{dU}{d\phi}+\frac{d\ln m(\phi)}{d \phi}\,\,(\rho-3p)\right].
\label{eomphiii} \ee
It is more convenient to replace one of the 
equations of motion by the 
conservation of the energy-momentum tensor. 
Combining the only non-trivial conservation equation with the equation of
motion (\ref{eomphii}) gives
\be
\frac{dp}{dr}=(3p-\rho)\,\frac{d\ln m(\phi)}{d \phi}\,\frac{d\phi}{dr}
-(\rho+p)\frac{1}{2B}\frac{dB}{dr}.
\label{solb} \ee

%\subsection{Additional expressions}

We also give some other useful forms of the Einstein 
equations.
Eq. (\ref{ein1}) can be integrated with the result
\be
A(r)=\left[1-\frac{1}{8\pi M^2} \frac{\calm(r)}{r} \right]^{-1},
\label{sola} \ee
where 
\be
\frac{d\calm(r)}{dr}=4\pi r^2
\left(\frac{1}{2A}\phi'^2+U(\phi)+\rho \right).
\label{mr} \ee
Eqs. (\ref{ein1}), (\ref{ein3}) can be combined in order to obtain
\be
\frac{d\ln (BA)}{dr}=\frac{1}{2M^2}\left[ 
\phi'^2+A(\rho+p)\right]r.
\label{solc} \ee

The Newtonian limit is taken easily if 
we combine eqs. (\ref{ein1})--(\ref{ein3}) in order to obtain
\begin{eqnarray}
B''+\frac{2}{r}B'&=&\frac{1}{4 M^2} AB \left[ 
(1+A)\rho + (5+A)p -4U +\left( \frac{1}{A}-1 \right) \phi'^2
\right]
\nonumber \\
&+&\frac{1}{8M^4}A B r^2 
\left[ A(p-U)(\rho+p)+\phi'^2
\left(\frac{1}{2}(\rho+3p)+
\frac{\phi'^2}{2A}-U  \right)
\right].
\label{bev0} \end{eqnarray}
For $B= 1+2\Phi$, with $|\Phi| \ll 1$, we can set $A=B=1$ in the r.h.s. 
of the above equation and neglect the second line.
In this way we find 
\be
\Phi''+\frac{2}{r}\Phi'=\frac{1}{4 M^2} \left( 
\rho + 3p -2U \right)
\label{newton} \ee
for the Newtonian potential. In the same limit, eq. (\ref{eomphiii})
becomes independent of the gravitational field and reads
\be
\phi''+\frac{2}{r} \phi'=
%\frac{\partial(U-p)}{\partial\phi}=
\frac{dU}{d\phi}+\frac{d\ln m(\phi)}{d \phi}\,\,(\rho-3p).
\label{newtonphi} \ee

\subsection{The Thomas-Fermi approximation}

The most popular models
of dark matter assume that it consists of a gas of weakly interacting
fermions. In the previous section we derived the formalism for our study
neglecting the quantum nature of the particles. As we are interested
in dense dark matter configurations, we must account for the fact 
that the quantum properties
may become relevant at sufficiently high density. 
In our treatment we shall only take into account the physical implications
of the exclusion principle. We neglect particle scattering,
but we assume that the gas is described by a Fermi-Dirac distribution. 
This is the essense of the Thomas-Fermi approximation.

In a local frame at every point in space the particles 
are described through the distribution 
\be 
f(p)=\left[\exp \left(
\frac{\sqrt{p^2+m^2(\phi(r))}-\mu(r)}{T(r)}
 \right) +1 \right]^{-1}.
\label{tf} \ee
The chemical potential $\mu$ and temperature $T$, 
as measured by a local observer,
are functions of the radial coordinate $r$, while the mass of the 
particles depends on the local value of the field $\phi$.
The pressure, number density and energy density are given by 
\begin{eqnarray}
p&=&T(r)\frac{1}{4\pi^3}\int d^3p\,\,
\ln \left[\exp \left(
-\frac{\sqrt{p^2+m^2(\phi(r))}-\mu(r)}{T(r)}
 \right) +1
\right]
\nonumber \\
~&=& \frac{1}{4\pi^3}\int d^3p\,\,f(p)\,\frac{p^2}{3\sqrt{p^2+m^2(\phi(r))}}
\label{pressure} \\
n&=& \frac{1}{4\pi^3}\int d^3p\,\,f(p)
\label{number} \\
\rho&=& \frac{1}{4\pi^3}\int d^3p\,\,f(p)\,\sqrt{p^2+m^2(\phi(r))}.
\label{energyd}
\end{eqnarray}
The above quantities are related through 
\be
p=\mu n -\rho + T s,
\label{grandpot} \ee 
where $s$ is the entropy density. 
An important identity is
\be
\frac{\partial p}{\partial \phi}=\frac{d\ln m(\phi)}{d \phi}\,\,(3p-\rho)
=\frac{d \ln m(\phi)}{d \phi}\,\,T^\mu_{~\mu}.
\label{ident} \ee

Combining this identity and the conservation equation (\ref{solb}) 
results in 
\be
\frac{\partial p}{\partial r}=-(\rho+p)\frac{1}{2B}\frac{dB}{dr}.
\label{soll} \ee
Evaluating the $r$-derivative of $p$, given by eq. (\ref{pressure}),
and employing the relation (\ref{grandpot}) permits us to rewrite
eq. (\ref{soll}) as
\be
Ts\frac{d\ln(T\sqrt{B})}{dr}+\mu n \frac{d\ln(\mu\sqrt{B})}{dr}=0.
\label{fin} \ee
This equation is satisfied if
\be
T(r)=T_0/\sqrt{B(r)},~~~~~~~~~~~~~\mu(r)=\mu_0/\sqrt{B(r)}.
\label{finn} \ee
These are the standard expressions that describe the behaviour of a
Thomas-Fermi fluid in a gravitational field.

The derivation of the equation of motion (\ref{eomphii}), on which
our discussion is based, neglected the
quantum nature of the particles. 
However, within the Thomas-Fermi approximation the only modification 
relative to a gas of classical particles is the change of the distribution.
When this is taken into account in the energy density and pressure, 
eq. (\ref{eomphii}) remains valid. For $T=0$, 
this has been shown explicitly in 
ref. \cite{tdlee1},
starting from the equation of motion of a fermionic field
(the Dirac equation) in a curved background.

A consistent solution can be obtained by integrating eqs.
(\ref{ein1}), (\ref{ein3}), (\ref{eomphii}) or (\ref{eomphiii}),
(\ref{solb}).
Eq. (\ref{ein2}) is then automatically satisfied. 
As we have shown, eq. (\ref{solb}) can be replaced by eqs. (\ref{finn}).
In this way, the system of equations to 
be solved is reduced to eqs. (\ref{ein1}), (\ref{ein3}), 
(\ref{eomphii}) or (\ref{eomphiii}).

\section{Solutions}
\label{known}
\setcounter{equation}{0}

\subsection{Compact astrophysical objects}

The Thomas-Fermi approximation in a gravitational
background has been considered repeatedly in the past. 
In refs. \cite{tdlee1,lynn} the formalism was developed for 
a degenerate fermionic gas with $T=0$, interacting with a scalar field
through the mass of the particles.
In ref. \cite{bilic} 
a fermionic gas with non-zero $T$ and $\mu$, but no scalar field,
was considered.
In the previous section we developed the formalism for 
the more general case of a fermionic
gas with non-zero $T$ and $\mu$, interacting with a scalar field 
through the mass of the particles. 

The set of equations we derived has various
solutions that depend on the assumptions made about the matter content and
the boundary conditions. We summarize briefly known solutions of 
eqs. (\ref{ein1}), (\ref{ein3}), (\ref{eomphii}) or (\ref{eomphiii}) 
with $T$, $\mu$ obeying eqs. (\ref{finn}). 

a) In the absence of a scalar field, a degenerate fermionic gas
at zero
temperature can form a compact astrophysical object, in which 
the gravitational attraction is balanced by the degeneracy pressure. 
Such solutions have been discussed in ref. \cite{bilic}.
The binding energy that keeps the star together is provided by 
gravity. For this reason these solutions are not stable for small
values of the total fermionic number. 
The mass is a decreasing
function of their radius for small masses. The 
possibility of non-zero temperature $T\ll \mu$ has also been considered.

b) If a scalar field with an appropriate potential is considered,
a degenerate fermionic gas 
can form astrophysical objects that
have been termed fermion soliton stars \cite{tdlee1} or fermion Q-stars
\cite{lynn}. These objects have $T=0$ and the fermions can be assumed to 
be ordinary nuclei. In particular examples, the scalar field 
has been identified either with
the $\sigma$-field associated with the nuclear forces, or with a mesonic
condensate ($\pi$- or $K$-condensate).  The main difference with the
previous class of solutions is that these objects are bound even in
the absence of gravity. 
The value of the scalar field that minimizes the fermionic mass does not
coincide with the mimimum of the potential. 
A new phase exists in which the fermions are essentially massless,
while the scalar field is displaced from its vacuum expectation value. 
An appropriate choice of the couplings guarantees that this phase
has positive binding energy. The gravitational interaction increases
the pressure at the center of the object, as well as the binding energy
per fermion.

The method for constructing solutions of this type
has been described in ref. \cite{lynn}. 
In the case that has been considered more often,
the fermionic mass is $m(\phi)=g\phi$. The chemical potential $\mu$ is
taken in a range such 
that the energy per fermion in the interior is smaller than
the mass of a free fermion. As a result, the fermions are trapped 
in the interior. 
A static, spherically symmetric solution of arbitrarily large radius
can be constructed by making use of eq. (\ref{ident})
in order to interpret eq. (\ref{newtonphi}) 
as the equation governing the motion of a particle in the potential
$U_{eff}=p-U$. This is possible through
the replacement $r\to t$ and $\phi\to r$.
The theory is assumed to have a vacuum characterized by a 
field value $\phi=\phi_1$ for which $U=0$ and $\rho=p=0$.
If the gravitational field is neglected, a new solution $\phi=\phi_0$
is needed, for which 
$\partial(p-U)/\partial \phi=0$ and $p-U=0$.
This can be achieved by adjusting the chemical potential.
If ${\partial(p-U)}/{\partial \phi}=-\ex$ at $\phi_0$ 
with $\ex \to 0+$,
any solution with $\phi'(r=0)=0$ takes a long ``time'' $r$ before
moving away from $\phi_0$ towards larger values.
At $r\simeq R$ the field starts ``rolling'' quickly
towards $\phi_1$. 
The ``friction'' term
$~\sim 2/r$ has a negligible effect, so that 
the ``rolling'' takes place between the two 
maxima of $U_{eff}$ near $\phi_0$ and $\phi_1$ without loss of energy. 
Soon after $r=R$ 
the fermionic density becomes zero, as $m(\phi)$ exceeds $\mu$. 
The interior of the configuration corresponds to $r\lta R$.

After the inclusion of gravity,
the solution in the interior is again determined by the 
condition ${\partial(p-U)}/{\partial \phi}=0$ \cite{lynn}.
The $r$-dependence of
$\mu$, according to eq. (\ref{finn}), implies that the field
$\phi$ now varies with $r$ as well. The variation is very slow, so that the
l.h.s. of eq. (\ref{eomphiii}) can be approximated by zero.
The quantity $p-U$ is non-zero in the interior of the solution
\cite{lynn}. Its value at the center must be adjusted so that 
the exterior solution, which involves only the scalar field,
is matched in the presence of gravity. 
The mass of these soliton stars is
an increasing function of the
radius for small masses.

c) Pure scalar field configurations, coupled to gravity,
have been discussed in ref. \cite{wetl}, where an explicit analytical solution
has been derived.
This solution, however, has
a naked singularity at the origin. 
The addition of fermions is necessary in order to eliminate the singularity.
In section \ref{astro} we present a solution in which this
is achieved.

\subsection{Dark matter halos}
%\label{halos}

The most prominent feature of the 
distribution of dark matter particles in
galaxy halos can be understood using the formalism of
section \ref{formalism}. The dark matter distribution results in approximately
flat rotation curves\footnote{
It must be emphasized that the rotation curves deduced from observations
or computer simulations are not exactly flat, but reveal a structure
related to the underlying physics of structure formation \cite{nfw}.
In particular, 
the distribution of matter in the galaxy cores is an open problem, for
which analytical approaches are not currently available. Theoretical
studies rely mainly on computer simulations. Here we make the assumption
that the approximate flatness of the rotation curves outside the galaxy cores
is related to the virialization of the dark matter gas. The simple model of
the isothermal sphere captures the essence of this assumption, while it can be
easily extended to include the scalar field. The solution of the Einstein
equations that we summarize is equivalent to the model of the isothermal
sphere. It cannot explain the details of the observed or simulated rotation
curves, apart from the gross feature of the approximate flatness outside
the cores. However,
it leads to important conclusions for the dispersion of the velocity of the 
dark matter particles, which are crucial for dark matter searches.
} 
for objects orbiting the galaxies at distances 
$r\gta$ 10 kpc \cite{peeblesb,silk}.
An analytical
understanding of this behaviour is possible within simple models of the
dark matter gas, such as 
the isothermal sphere. If these simple models are extended
through the addition of a scalar field to the
theory, an analytical treatment is still feasible.
For our study we model the dark matter gas as a Thomas-Fermi fluid, described
by the formalism of section \ref{formalism}.
We assume that the gas is non-relativistic and 
non-degenerate,
with $p \ll \rho$. 
In the absence of a scalar field ($\phi=0$, $U=0$, $m=m_0$)
we can define the non-relativistic chemical potential
as $\mub_0=\mu_0-m_0$. 
The number density is 
\be
n\simeq 2\left( \frac{m_0T}{2\pi}\right)^{3/2} 
\exp \left(-\frac{m_0}{T}+\frac{\mu}{T} \right)
\simeq 
n_0 \exp \left(-\alpha \Phi \right),
\label{num1} \ee
with
\be 
n_0=2\left( \frac{m_0T_0}{2\pi}\right)^{3/2} 
\exp \left(\frac{\mub_0}{T_0} \right),
~~~~~~~~~~~~~
\alpha=\frac{m_0}{T_0}. 
\label{ca} \ee
The validity of the above expressions requires
$\mub_0 <0$ and $T_0\ll m_0$, $T_0\ll |\mub_0|$. 
The gravitational field can be studied in the Newtonian limit.
Then, eq. (\ref{newton}) can be written as
\be
\frac{d^2u}{dz^2}+\frac{2}{z}\frac{du}{dz}+\exp u=0,
\label{uev} \ee
with $u=-\alpha \Phi$, $z=\beta r$, $\beta^2=\alpha \rho_0/4M^2$, and
$\rho_0\simeq m_0 n_0$.
The solutions of this equation that are regular at $z=0$ 
%(we assume $u=0$, $du/dz=0$) 
behave as $\exp u=2/z^2$ for large $z$ 
\cite{peeblesb}. 
For large $r$ the Newtonian potential 
varies only logarithmically with $r$, 
while the integrated mass of the dark matter scales
linearly with $r$. This leads to flat rotation curves for objects
orbiting the galaxy \cite{peeblesb}.

In ref. \cite{hhalos} the possibility was considered 
that the mass of the dark matter
particles depends on the field $\phi$ that 
varies slowly with the radial distance $r$. The analysis can be easily
reproduced through the formalism we developed in section \ref{formalism}.
We approximate the number density of 
the dark matter particles by the 
expression (\ref{num1}), with $m_0$ replaced by $m(\phi)$. 
As before, we work in the weak-field limit, in which
$B(r)\simeq 1+2\Phi(r)$, with $|\Phi(r)|\ll 1$.
We assume that the field $\phi$ is displaced from
its asymptotic  ($r\to \infty$) value only by a small amount, 
so that the approximation 
$m(\phi)=m(\phi_0)+[dm(\phi_0)/d\phi]\, \dphi
\equiv m_0+m'_0 \, \dphi$ can be employed.
We identify $\phi_0$ as the value of the field at 
the center of the galaxy ($r=0$). 
We work in the leading order in $\dphi$, and
assume that $m'/m\simeq m'_0/m_0$ for all $r$.
Our treatment is relevant up to 
a distance $r_1 \sim$ 100--200 kpc beyond which the dark matter becomes very
dilute. For $r \gta r_1$, 
we expect 
that $\phi$ quickly becomes constant with a value close 
to $\phi(r_1)\equiv \phi_1$. 
This is the value that
drives the present cosmological expansion
\cite{wetdil,peeblesold,wetcosmon}. Here we assume that the
cosmic evolution of 
$\phi_1$ is negligible for the time-scales of interest, so that
the asymptotic configuration is static to a good approximation. 

Within the leading order in $\dphi$, we can approximate
$dU/d\phi$ by a constant between $r=0$ and $r=\infty$.
For the scalar field to provide a resolution of the coincidence problem,
the two terms in the r.h.s. of eq. (\ref{eomphi}) must be of 
similar magnitude in the cosmological solution.
This means that 
$dU/d\phi$ must be comparable to $(m'_0/m_0)\rho_\infty$.
We expect $\rho_\infty$ to be a fraction of the critical density, i.e. 
$\rho_\infty \sim 3$ keV/cm$^3$. On the other hand, for the spherically
symmetric solution the energy density in the r.h.s. of
eq. (\ref{eomphi}) is that of the galaxy halo 
($\sim 0.4$ GeV/cm$^3$ for our neighborhood of the Milky Way).
This makes $dU/d\phi$ negligible in the r.h.s. of
eq. (\ref{eomphi}) for a static configuration. The potential is
expected to become important only for $r \to \infty$, where the
static solution must be replaced by the cosmological one.
Similar arguments indicate that we
can neglect $U$ relative to $\rho$. 
The scale for the field mass is expected to be
set by the present value of the Hubble parameter. Then
the field is effectively massless at distances of 
the order of the galactic scale. 
The above indicate that, if the deviation 
of the scalar field from its cosmological value is small, the form of the 
potential plays a negligible role at the galactic level.
For this reason we set $U=0$ in our discussion. 

The scalar field 
generates a new long-range scalar interaction, whose strength relative
to the gravitational interaction is 
determined by the parameter 
\be
\kappa^2=4M^2\left( \frac{m'_0}{m_0}\right)^2.
\label{kappa} \ee 
If the new interaction is universal for ordinary and dark matter, the
experimental constraints impose $\kappa \ll 1$. In this case, it is
reasonable to expect a negligible effect in the distribution of
matter in galaxy halos. 
However, if $\phi$ interacts only with dark matter this constraint can 
be relaxed. In the following we assume that the scalar field 
interacts only with the dark matter.

As before, we assume
$p\ll \rho$ and employ the nonrelativistic chemical potential
$\mub_0=\mu_0-m_0$. 
The number density of dark matter can be written as 
\be
n\simeq 2\left( \frac{m(\phi)T}{2\pi}\right)^{3/2} 
\exp \left(-\frac{m(\phi)}{T}+\frac{\mu}{T} \right)
\simeq 
n_0 \exp \left(-\alpha \Phi-\at \dphi \right),
\label{num11} \ee
with $n_0$, $\alpha$ given by eqs. (\ref{ca}) and 
\be
\dphi=\phi-\phi_0=\phi-\phi(r=0),~~~~~~~~~~~~~~~~~\at=\frac{m'_0}{T_0}. 
\label{att} \ee
The energy density of dark matter at the center of the galaxy is 
$\rho_0=m_0 n_0$.

It is important to emphasize that the assumption that the number density
is given by eq. (\ref{num1}) does not require the presence of thermal
equilibrium. 
In an alternative approach, followed in ref. \cite{hhalos},
the dark matter can be considered as a dilute, weakly interacting
gas with an
energy-momentum tensor 
$T^\mu_{~\nu}={\rm diag} (-\rho,p,p,p)$.
Motivated by the model of the isothermal sphere \cite{peeblesb},
we can assume that 
$p(r)=\rho(r)\, \langle \vrm_d^2 \rangle
=m(\phi(r))\, n(r)\, \langle \vrm_d^2 \rangle$, with a constant 
velocity dispersion $\langle \vrm_d^2 \rangle \ll 1$. 
In the weak field limit and for
$p \ll \rho$, the conservation of the energy-momentum tensor (\ref{solb})
gives 
\be
p'=-\rho\, \Phi'-\rho \, \frac{m'_0}{m_0} (\dphi)'.
\label{extra3} \ee
Integration of this equation results in
\be
n\simeq n_0
\exp \left(-\frac{\Phi}{\langle \vrm_d^2 \rangle} 
-\frac{m'_0}{m_0}\frac{\dphi}{\langle \vrm_d^2 \rangle} \right).
\label{extra4} \ee
By defining an effective temperature $T_0$
through the relation
$\langle \vrm_d^2 \rangle=T_0/m_0$, we reproduce eq. (\ref{num11}).
This indicates that the assumption of thermal equilibrium is not 
required for the emergence of eq. (\ref{num11}). The parameter $T_0$ 
that appears in the various expressions of
section \ref{formalism} 
does not correspond necessarily to the physical temperature.
In many cases of physical interest it is simply a measure of the
typical velocity of the Fermi gas. 

In the non-relativistic, weak-field limit, with $p=U=0$, 
eq. (\ref{newton}) becomes
\be 
\Phi''+ \frac{2}{r}\Phi'=\frac{1}{4M^2} \, \rho_0 
\exp \left(-\alpha \Phi-\at \dphi \right).
%+\frac{1}{4M^2} \, \rho_B \,\,
%f \left(\frac{r}{r_c} \right).
\label{eoma} \ee
Similarly , eq. (\ref{newtonphi}) becomes
\be
(\dphi)''+\frac{2}{r}(\dphi)'=\frac{m'_0}{m_0}\, \rho_0 
\exp \left(-\alpha \Phi-\at \dphi \right).
\label{eomb} \ee
A linear combination of eqs. (\ref{eoma}), (\ref{eomb}) results in
eq. (\ref{uev})
where now
\be
u=-\alpha \Phi-\at \dphi,~~~~~~~~z=\beta r,
\label{udef} \ee
with 
\be
\beta^2=\left( \frac{\alpha}{4M^2}+\at \frac{m'_0}{m_0}\right) \rho_0
=(1+\kappa^2)\frac{m_0}{T_0}\frac{\rho_0}{4M^2}. 
\label{beta} \ee
The solutions that are regular at $z=0$ approach the form
\be
u=\ln\left(\frac{2}{z^2}\right)+\frac{1}{\sqrt{z}}
\left[
d_1\cos\left(\frac{\sqrt{7}}{2}\ln z \right)
+d_2\sin\left(\frac{\sqrt{7}}{2}\ln z \right)
\right]+...
\label{ucor} \ee
for large $z$.
Another linear combination of eqs. (\ref{eoma}), (\ref{eomb}) gives 
\be 
\frac{d^2v}{dz^2}+\frac{2}{z}\frac{dv}{dz}
=0,
\label{eomc} \ee 
with 
\be
v=\at \dphi-4M^2\frac{m'_0}{m_0}\at\Phi=-\kappa^2 \alpha \Phi+\at\dphi.
\label{vdef} \ee
The solution of this equation is 
$v=c_0+{c_1}/{z}.$

The velocity $\vrm$ of a massive baryonic object in orbit around the galaxy, 
at a distance $r$ from
its center, can be expressed as 
\be
\left(\frac{\vrm}{\vrm_c}\right)^2=\frac{r\Phi'}{\vrm^2_c}=-\frac{z}{2}
\left(\frac{du}{dz}+\frac{dv}{dz} \right),
\label{rotvel} \ee
where
\be
\vrm_c^2=\frac{2}{1+\kappa^2}\frac{T_0}{m_0}=\frac{2}{1+\kappa^2}
\langle \vrm^2_{d} \rangle.
\label{vel} \ee
The asymptotic form of $u(z)$, $v(z)$
indicates that $\vrm \simeq \vrm_c$ for large $z$. The
dominant correction to the leading behaviour 
arises from the term $\sim 1/\sqrt{z}$ in eq.
(\ref{ucor}). The function $v(z)$ gives a higher order correction.
This means that the presence of the 
field $\phi$ is not expected to cause significant modifications
to the shape of the rotation curves 
relative to the $\phi=0$ case.
For a vanishing field the rotation curves are 
again governed by the solution of eq. (\ref{uev}), given by eq. (\ref{ucor}).
This simple analysis indicates that 
the approximately flat rotation curves outside the galaxy cores
are a persistent feature 
even if the dark matter is coupled to a scalar field through
its mass. However, numerical simulations are probably necessary
in order to reproduce 
the detailed form of the curves.

We can use eq. (\ref{vel}) in order to fix 
$T_0/m_0=\langle \vrm^2_{d} \rangle$ for
a given value of $\kappa$. 
The effect of the scalar field is encoded in the factor
$\kappa^2=4M^2(m'_0/m_0)^2$. When this is small, the velocity of
an object orbiting the galaxy is of the order of the square root of
the dispersion of the 
velocity of the dark matter particles. If $\kappa^2$ is large the 
orbital velocity can become much smaller than the typical
dark matter velocity. 
This behaviour persists even if baryonic
matter is added near the center of the galaxy \cite{hhalos}.
Within our extension of 
the model of the isothermal sphere, the flatness of the rotation curves
is largely insensitive to the details of the matter distribution within
the galaxy core.

The scenario we considered may be more interesting for dark 
matter searches than the conventional one. For large values of $\kappa$
the velocity of dark matter particles exceeds significantly the
observed rotation velocity ($\sim 220$ km/s for the Milky Way). 
The estimated local energy density of dark matter remains the same as
in the case with $\kappa^2=0$.
It is $\sim 0.4$ GeV/cm$^3$ for our neighborhood of the Milky Way. 
As a result the flux of dark matter particles towards a terrestrial
detector is larger roughly by a factor $(1+\kappa^2)^{1/2}$ relative to the 
$\kappa^2=0$ case.

A detailed calculation of the counting rates in detectors must take into 
account the motion of the Earth around the Sun and 
the motion of the Sun through our galaxy \cite{factors}. 
The velocity distribution of dark matter is
\be
f(\vec{\vrm}) \sim \exp\left(  
-\frac{\left| \vec{\vrm}+\vec{\vrm}_E+\vec{\vrm}_S \right|^2}{
2\langle \vrm^2_d \rangle}
\right),
\label{veldistr} \ee
where 
the magnitude of the velocity of the Earth relative to the Sun is 
$|\vec{\vrm}_E|\simeq$ 30 km/s, and that of the 
Sun relative to the galactic rest frame $|\vec{\vrm}_S|\simeq
\vrm_c\simeq$ 220 km/s.
For $\kappa \gta 1 $, we have $\sqrt{\langle\vrm^2_d \rangle} \gg
|\vec{\vrm}_E|,|\vec{\vrm}_S|$. The
motion of the Earth and the Sun are
expected to give only a small modification of the 
dark matter flux towards the Earth. As a 
result, the seasonal variation of a possible dark matter signal 
decreases for increasing $\kappa^2$.
Typically, 
the cross section for the elastic scattering of halo particles by target
nuclei is independent of the
particle velocity for very low velocities \cite{witten,silk}.
The leading effect of a non-zero value of $\kappa$ 
is that the counting rates, that are
proportional to the dark matter velocity, are increased
by the factor $(1+\kappa^2)^{1/2}$. 
This makes the dark matter easier to detect. 
Existing bounds on dark matter properties from direct searches
can be extended to include the case of non-zero $\kappa$.
The bound on the cross section for the interaction  
of dark matter with the material of the detector must be strengthened by 
the factor $(1+\kappa^2)^{1/2}$. 

Corrections to the above simple picture may arise for large values of
$\kappa^2$. For example, the nuclear form factors that must
be included in the calculation of the counting rates have a 
velocity dependence \cite{silk}.
If the velocities are larger than the conventionally assumed 
$\sqrt{\langle\vrm^2_d \rangle}\sim 220/\sqrt{2}$ km/s, these form factors
may result in significant modifications of the expected rates for the
various experiments. A separate detailed study is necessary in
order to address this issue.

The allowed range of $\kappa$ is limited by the observable implications
of the model that describes the dark sector. The dependence of the mass of 
dark matter particles on an evolving scalar field 
during the cosmological evolution since the 
decoupling is reflected in the microwave background. The magnitude of the
effect is strongly model dependent. In the models of ref. 
\cite{amendola,mainini} the observations result in
the constraint $\kappa^2 \lta 0.01$. 
In the model of refs. \cite{peeblesfar,peeblessim} the 
scalar interaction among dark matter particles is screened by an additional
relativistic dark matter species. As a result, 
the model is viable even for couplings 
$\kappa^2 \simeq 1$. 
A similar mechanism is employed in ref. \cite{massimo}.
In this model
the interaction between dark matter and dark energy 
becomes important only during the recent evolution
of the Universe. In general, an interaction that is effective for 
redshifts $z \lta 1-2$ is not strongly constrained by the 
observations.

Before concluding this section we would like to give a more physical
interpretation of the main physical effect. 
The dark matter in galaxy halos is expected to be a virialized gas, whose
velocity is determined by the potential. The presence of the
field associated with the dark energy
adds an attractive scalar potential to the gravitational one.
Their relative strength is given 
by the parameter $\kappa^2$ defined in eq. (\ref{kappa}).
As the scalar field is essentially massless, both potentials have
the same radial dependence.  
Their combined effect results in a total potential that is stronger
by a factor $1+\kappa^2$ compared to the conventional case. 
As a result, the kinetic energy of the virialized gas 
and the velocity dispersion are inreased by
the same factor.

\section{New astrophysical objects}
\label{astro}
\setcounter{equation}{0}

\subsection{The ingredients for a solution}

In the previous section we discussed the effect of the scalar field on
the distribution of dark matter in galaxy halos. Our basic assumption 
was that the field is not displaced significantly from its value 
at $r\to \infty$. 
%We found that, 
%for a dilute gas of dark matter particles that
%follows an approximate Maxwell distribution, even 
%a small shift of the field value can alter  
%the distribution significantly, 
%as the effect on the dark matter mass appears in the
%exponent. 
In this section we are interested in solutions with 
larger deviations of 
the field from its asymptotic
value. For this reason we need a 
treatment that goes beyond linearized gravity. 

In the framework we are considering, the potential $U(\phi)$
does not have a minimum. The typical case, which we shall
take as a working example, is an exponential potential 
$U(\phi)=U_0 \exp(-c\phi/M)$
\cite{wetdil,wetcosmon,amendola}.
There are two possibilities, depending on the form of 
$m(\phi)$:
\\
a) If the mass $m$ is a decreasing function of $\phi$
(such as the decreasing 
exponential of refs. \cite{wetcosmon,amendola}), the pressure
$p$ becomes an increasing function of $\phi$.
This is obvious from eq. (\ref{ident}), taking into account that 
$p\leq \rho/3$. 
For a weak gravitational field
we can make use of eq. (\ref{ident})
in order to interpret eq. (\ref{newtonphi}) 
as the equation governing the motion of a particle in the potential
$U_{eff}=p-U$. This is possible through
the replacement $r\to t$ and $\phi\to r$.
The potential $U_{eff}$ is an increasing function of 
$\phi$. 
Any static configuration must start with a large value of $\phi$ at
$r=0$,
which then ``rolls'' for $r\to \infty$ to a 
smaller value $\phi_1$ that is relevant for
the cosmological solution.  
In more physical terms, this behaviour can be understood through the
tendency of the system to minimize the energy by reducing the
particle mass. As this mass vanishes for $\phi\to\infty$, we expect that
the field becomes larger than its cosmological value 
in the interior of the most stable static configuration. 
If we assume that during the  
cosmological evolution (which we neglect in the study of static 
configurations) the field
$\phi$ moves from smaller to larger values, such
configurations may emerge dynamically if the 
field in certain regions grows faster than
in the bulk of space.
A strong gravitational field is not expected to induce qualitative
modifications of this behaviour.
\\
b) There are also scenaria in which the dark matter mass is an
increasing function of $\phi$. For example, in refs. 
\cite{peeblesfar,peeblessim} the mass of the fermions that compose the dark
matter is determined by the expectation value of the scalar field 
through a Yukawa coupling. We do not consider all the details 
of the scenario of refs. 
\cite{peeblesfar,peeblessim}, which involves two dark matter families with
different masses. 
Instead, we limit ourselves to one family of dark matter particles.
If the mass is an increasing function of the field,
it is reasonable to imagine that, during the
evolution of the field from smaller to larger values, certain regions
of space become disconnected from the general evolution because dark
matter gets trapped there. The field remains close to zero in these
regions. The vanishing of the dark matter mass for 
zero $\phi$ makes such a scenario energetically favourable. 

In the examples we shall consider we shall assume that the dark matter
mass is linear in $\phi$. The essence of our assumption is that the mass
vanishes for some finite value of $\phi$, which can be
set to zero by an appopriate field shift. The potential $U(\phi)$ retains
its form, as the field shift 
can be absorbed in the pre-exponential factor.
The strong dependence of the mass on $\phi$ causes the field near dense 
concentrations of dark matter to 
deviate strongly from its asymptotic value $\phi_1$. The field
tends to approach zero in the center of such concentrations, so that
the dark matter particles become massless there.
Our framework corresponds to the second case discussed above. 
However, the same physical mechanism could also operate in the 
first case, for which the field is expected to
be larger than $\phi_1$ in the interior of concentrations of dark matter. 
Astrophysical configurations similar to the ones we 
describe in the following are
expected to appear in this case as well. 

The solutions we shall derive are very different from the 
solutions of the previous section, as the dark matter gas will be 
degenerate ($T=0$).
It is possible to find solutions with $T\not= 0$ following an analogous
procedure, even though the technical difficulty may be greater. 

We have seen that 
the effective temperature of the fermionic gas can be related either to the 
velocity dispersion of a weakly interacting 
collection of particles (as for galaxy
halos), or to the real temperature of a thermalized gas (as for the interior of
stars). Our formalism can describe static solutions in 
both cases. The value of the temperature
is determined by the ability of a collapsing gas to lose momentum, 
and the possible interactions between the particles that can
establish thermalization. We shall not address 
this dynamical issue here. 
In the remaining of this section 
we shall present static solutions that describe
a system very different from the galaxy halos we discussed in the 
previous section: a 
completely degenerate fermionic gas, interacting with a scalar field.

The pure scalar field configuration
that solves the Einstein equations in the absence of a potential and 
dark matter has been discussed in ref. 
\cite{wetl}. We expect this solution to be
realized for distances $r>R$, for which
the dark matter density becomes negligible (of the order of the critical
density of the Universe). The dark matter gas is concentrated in the region 
$r\leq R$. The radius of the compact object is $R$. 
For $U=0$ and large $r$, the leading terms of the pure field solution read
\be
\phi=\phi_1-\gamma M \frac{R_s}{r}
~~~~~~~~~~~~~~~
B=A^{-1}=1-\frac{R_s}{r}.
\label{asympt} \ee
The free parameter $R_s$ determines the mass of the object as seen from
an observer at infinity
\be
M_{tot}=8\pi M^2 R_s.
\label{mass} \ee
The asymptotic value of the field for $r\to \infty$ is $\phi_1$, while
the free parameter $\gamma$ determines the derivative energy density 
\be
\calk=\frac{1}{2A}\phi'^2=\frac{\gamma^2 R^2_s M^2}{2r^4}.
\label{gradient} \ee
As long as $\calk\gg U$, the potential can be neglected. For distances
$r$ of the order of the Hubble radius this condition is not expected to
be satisfied. At such scales the solution we are describing must be
replaced by the time-dependent cosmological solution.

\subsection{The equations}

We would like to modify the solution of ref. \cite{wetl} by
adding dark matter in the region $r\leq R$.
This addition can eliminate naturally the
singularity that appears in the pure field configuration.
The simplest form of the  mass of the dark matter particles
is 
\be
m(\phi)=m_0+g \phi,
\label{mass1} \ee
where we have assumed a Yukawa interaction between the dark matter fermions
and the scalar field. The mass $m_0$ is expected to be associated with
a new scale $\sx$, different from the Planck scale $M$. 
For example, $\sx$ could be set by the expectation value of a heavy Higgs
field in a Grant Unified Theory. In order for the Yukawa interaction not to
generate a contribution to the mass much larger than $m_0$, the Yukawa 
coupling must be taken $g \lta \sx/M$.  
Then the constant $m_0$ can be absorbed in a shift of $\phi$ of order $M$. 
In the following we shall assume that the mass is 
\be
m(\phi)=\sx \frac{\phi}{M},
\label{massex} \ee
where we have taken $g=\sx/M$ for simplicity.
It must be pointed out that our assumption does not provide an explanation
for the smallness of the Yukawa coupling. The appearance of a small parameter
is inevitable in a system with two physical scales: the expectation value
of $\phi$, taken of order $M$, and the physical mass of the dark matter 
particles, often assumed to be as low as the TeV range. The situation is
similar to the flavour problem in the Standard Model, for which a deeper
understanding is lacking.  

We assume that the scalar field has a potential of the form
\be
U(\phi)=C \sx^4\exp\left(-c\frac{\phi}{M} \right).
\label{potex} \ee
with $c={\cal O}(1)$. 
The value of $\phi$ that is relevant for the present cosmological evolution
is given by the requirement that $U(\phi)$ be of the order of the 
critical energy density $U(\phi)\sim 10^{-47}$ GeV$^{4}$. This value
\be
\frac{\phi_1}{M}=\phit_1 \simeq \frac{1}{c}
\left[ 108+\ln C +4 \ln \left( \frac{\sx}{{\rm GeV}}\right)\right]
\label{valueasy} \ee
must be approached by our static solution for $r\to \infty$.
A possible constant contribution $m_*$ to the mass defined in 
eq. (\ref{massex})
has been absorbed in $\phi$. Through this redefinition 
$\phi/M \to \phi/M+m_*/\sx$.
We assume implicitly that $m_*=\cal{O}(\sx)$ in order to avoid
unnaturally large values of $\phi$.

The equations of motion become more transparent if we define the
dimensionless variables 
\be
\phit=\frac{\phi}{M},
~~~~~~~~~~~~~~~~~
\rt= \frac{\sx^2r}{M}.
\label{dimension} \ee
All other dimensionful quantities are multiplied with appropriate
powers of $\sx$ only, in order to form dimensionless quantities denoted as
tilded. We also define the quantity 
\be
\Bt = \frac{B}{\mut^2_0}=\frac{B\sx^2}{\mu_0^2}.
\label{betatild} \ee
We have the relations 
\be
\mt(\phit)= \phit,
~~~~~~~~~~~~~~~~~
\mut(\rt)=\frac{1}{\sqrt{\Bt(\rt)}}.
\label{aa2} \ee
We define the surface of the compact object as the
point at which the fermionic density and pressure vanish. 
If the surface corresponds to a value
$\rt=\Rt={\cal O}(1)$, the physical radius is given by the 
relation
\be
\frac{R}{{\rm km}}\simeq 0.34\left(\frac{{\rm GeV}}{\sx}\right)^2 \Rt.
\label{physr} \ee
Similarly, from eq. (\ref{mass}), we find
\be
\frac{M_{tot}}{M_\odot}\simeq 0.115 \left(\frac{{\rm GeV}}{\sx}\right)^2 \Rt_s.
\label{massr} \ee
%In general, we expect $\Rt$ and $\Rt_s$ to be of the same order of 
%magnitude. 
Another important characteristic of the solution is the total 
fermionic number, which we assume to be conserved. The fermionic
number density $j^0$ is the time component of a covariant 4-vector. We can
deduce its value from the local density $n$, through the 
appropriate tetrad factor $V^a_{~\mu}$, where 
$g_{\mu\nu}=V^a_{~\mu}V^b_{~\nu}\eta_{ab}$.
The total fermionic number is
\be
N=\int \sqrt{-g}\,d^3x \,j^0=
\int_0^\infty 4 \pi r^2 \,dr \sqrt{A} \,n 
=\left(\frac{M}{\sx} \right)^3 \int_0^\infty 4 \pi \rt^2 \,d\rt \sqrt{A} \,\nt
=\left(\frac{M}{\sx} \right)^3\Nt. 
\label{fermionum} \ee

The equations of motion become
\begin{eqnarray}
\phit''+\left[\frac{1+A}{\rt}-\frac{1}{2}A\rt\left( \Ut(\phit)
+\frac{1}{2}(\rhot-\pt) 
\right) \right] \phit'&=&
A\left[ \frac{d\Ut}{d\phit}+\frac{1}{\phit}\,\,(\rhot-3\pt)\right],
\label{eq1d} \\
\frac{1}{\rt^2}\frac{1}{A}-\frac{1}{\rt^2}-\frac{1}{\rt}\frac{A'}{A^2}
&=&\frac{1}{2}\left( -\frac{1}{2A}\phit'^2-\Ut(\phit)-\rhot\right),
\label{eq2d} \\
\frac{1}{\rt^2}\frac{1}{A}-\frac{1}{\rt^2}+\frac{1}{\rt}
\frac{\Bt'}{\Bt A}
&=&\frac{1}{2}\left( \frac{1}{2A}\phit'^2-\Ut(\phit)+\pt\right),
\label{eq3d} 
\end{eqnarray}
where a prime denotes a derivative with respect to $\rt$.
We also have
\begin{eqnarray}
\nt&=&\frac{1}{3\pi^2} \left(\mut^2-\mt^2  \right)^{3/2},
\label{nd} \\
\pt&=&\frac{1}{24\pi^2}
\left[\mut \sqrt{\mut^2-\mt^2}\left(2\mut^2-5\mt^2 \right)
+3\mt^4\ln\left(
\frac{\mut+\sqrt{\mut^2-\mt^2}}{\mt} \right)
\right],
\label{pd} \\
\rhot&=&\frac{1}{8\pi^2}
\left[\mut \sqrt{\mut^2-\mt^2}\left(2\mut^2-\mt^2 \right)
-\mt^4\ln\left(
\frac{\mut+\sqrt{\mut^2-\mt^2}}{\mt} \right)
\right],
\label{rhod} 
\end{eqnarray}
for $\mut\geq \mt$, and $\nt=\rhot=\pt=0$ for $\mut< \mt$.
Finally,
\be
\Ut(\phit)=C \exp\left(-c \phit \right).
\label{utt} \ee

We need four 
initial conditions for the system of equations (\ref{eq1d})--(\ref{eq3d}).
One of them is imposed by the regularity of the spherically
symmetric solution at $\rt=0$: $\phit'(0)=0$.
Another one is implied by eqs. (\ref{sola}), (\ref{mr}): $A(0)=1$.
The value of $\Bt(0)$ can be chosen arbitrarily. However, the normalization of
eqs. (\ref{asympt})
implies that $AB(r\to\infty)=1$.
This means that $A\Bt(\rt\to\infty)=(\mu_0/\sx)^{-2}$, where we have
used the definition (\ref{betatild}). As a result, the choice of
$\Bt(0)$ determines the chemical potential.
Finally, $\phit(0)$ must be chosen so that 
$\phit(\rt\to\infty)$ reproduces correctly the present
value $\phit_1$ of the scalar field in the cosmological
solution, as given by eq. (\ref{valueasy}). 
(We assume that the time scale of the cosmological 
solution is very large and neglect the time dependence of 
$\phi(r\to\infty)$.)

\begin{figure}[t]
 %\vspace{1.cm}
 \centerline{\epsfig{figure=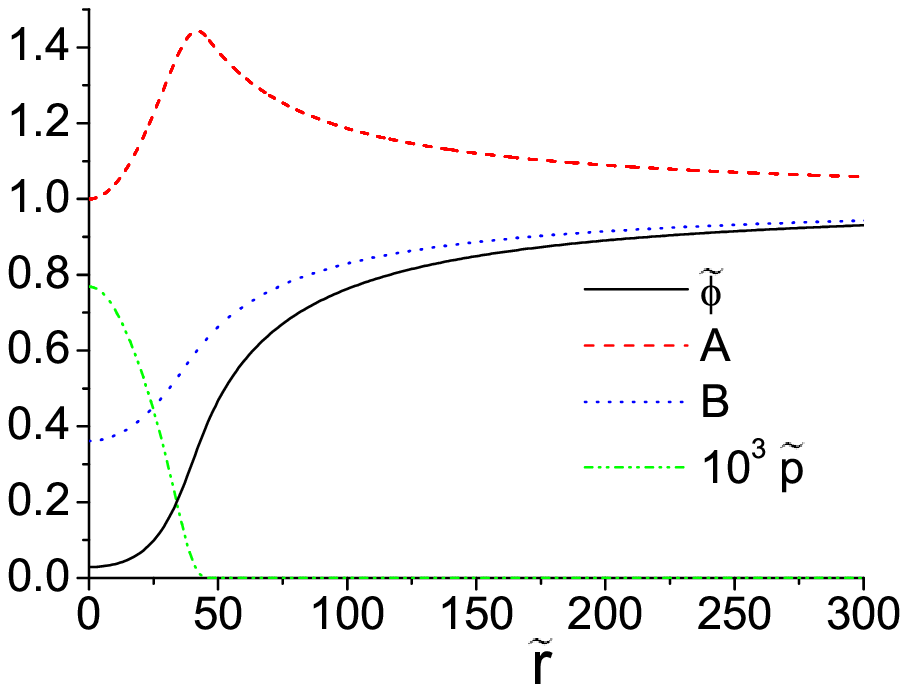,width=15cm,angle=0}}
 \caption{\it
The solution of eqs.
(\ref{eq1d}), (\ref{eq2d}), (\ref{eq3d}) 
for a model with $\sx/{\rm GeV}=1$, $\mu_0/\sx\simeq 0.33$, $\phi_1/M=1$,
$c=1$, $C\simeq 3.4 \times 10^{-47}$. We
plot $\phit$, $\pt$, $A$, $B$,
as a function of $\rt$.
The object has $\Rt\simeq 46$, $\Rt_s\simeq 17$, $\Nt \simeq 
920$.
}
 \label{fig3}
 \end{figure}

\subsection{The solution}

In fig. \ref{fig3} we display the form of the solution 
for a model with $\sx/{\rm GeV}=1$.
The chemical potential takes the value $\mu_0/\sx\simeq 0.33$, while
the scalar field approaches the value $\phi_1/M=1$ for large $r$.
The potential $U(\phi)$ has $c=1$, $C\sim 3 \times 10^{-47}$ and
is negligible in the range of distances of interest. 
We plot the quantities $\phit$, $\pt$, $A$, $B$
as functions of $\rt$. We observe that the scalar field approaches zero
near the center of the solution, so that the fermions become 
almost massless there. 
The pressure and density of the fermionic gas 
vanish for $\rt \geq \Rt \simeq 46$. 
The value of $\Rt$ determines the radius of
the astrophysical object through eq. (\ref{physr}):
$R \simeq 16$ km.
The metric components $A$ and $B$ deviate significantly from 1 in the interior
of the solution. For this reason a fully non-linear treatment of the 
Einstein equations has been necessary. The mass of the object can be
deduced from the asymptotic form of $A$ or $B$ for $\rt \to \infty$ through
the second of eqs. (\ref{asympt}). We find $\Rt_s \simeq 17$, which 
corresponds through 
eq. (\ref{massr}) to $M_{tot} \simeq 2.0 \, M_\odot$.
The total fermionic number is $\Nt\sim 920$, which gives
$N \simeq 4.7 \times 10^{57}$ through
eq. (\ref{fermionum}).

\begin{figure}[t]
 %\vspace{1.cm}
 \centerline{\epsfig{figure=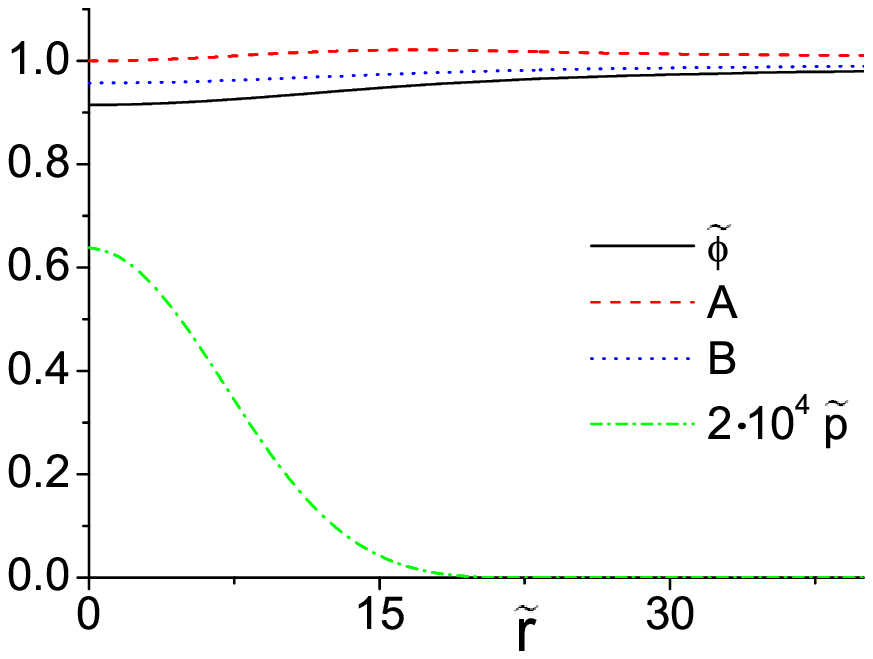,width=15cm,angle=0}}
 \caption{\it
The solution of eqs.
(\ref{eq1d}), (\ref{eq2d}), (\ref{eq3d}) 
for a model with $\sx/{\rm GeV}=1$, $\mu_0/\sx\simeq 0.95$, $\phi_1/M=1$,
$c=1$, $C\simeq 3.4 \times 10^{-47}$.
We plot $\phit$, $\pt$, $A$, $B$,
as a function of $\rt$.
The object has $\Rt\simeq 22$, $\Rt_s\simeq 0.41$,
$\Nt \simeq 10.5$.
}
 \label{fig4}
 \end{figure}

The potential $U(\phi)$ does not play any role in the solution we presented.
The reason is that its value is much smaller than 
the pressure or the field derivative energy. The field
essentially ``rolls'' on the effective potential 
$U_{eff}=p$.
Only for $\rt \to \infty$ we 
expect $U$ to become important. In this limit, our solution must be
replaced by the time dependent cosmological solution.

\begin{figure}[t]
 %\vspace{1.cm}
 \centerline{\epsfig{figure=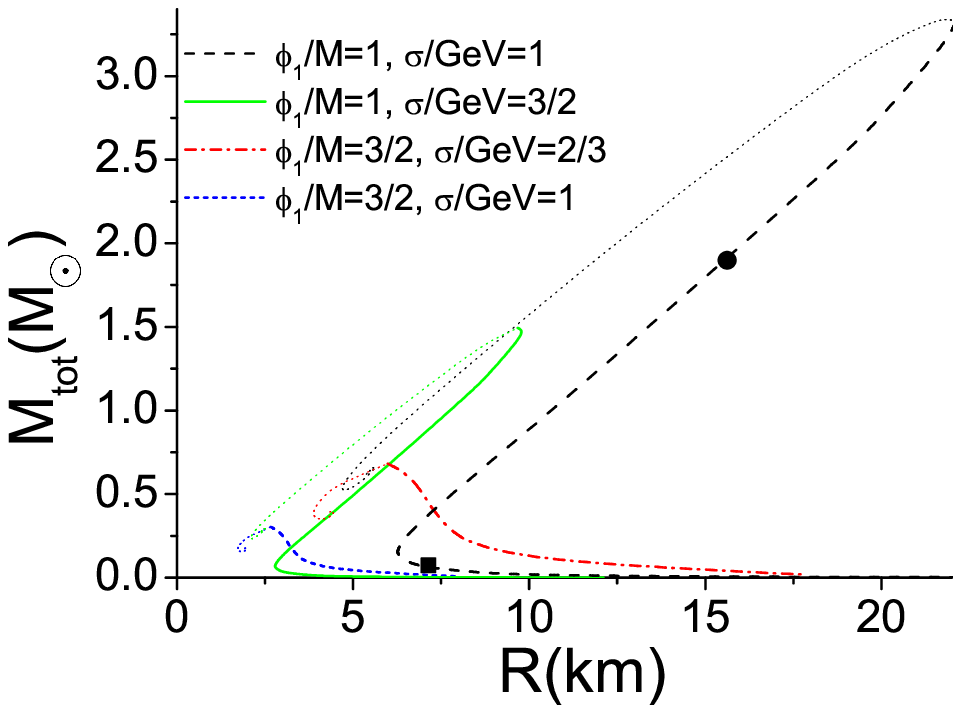,width=15cm,angle=0}}
 \caption{\it
The function $\Rt_s(\Rt)$ for 
various models. }
 \label{fig5}
 \end{figure}

In fig. \ref{fig4} we present the form of the solution if
the chemical potential is $\mu_0/\sx\simeq 0.95$ while the
remaining parameters remain the same. The deviations of $\phit$, $A$ and $B$
from 1 are much smaller than in the previous case. The fermionic gas is
much more dilute, as its pressure is approximately 20 times smaller.
The resulting astrophysical object has 
$\Rt\simeq 22$, $\Rt_s\simeq 0.41$, which
correspond to $R \simeq 7.4$ km, $M_{tot} \simeq 0.047 M_\odot$.
The total fermionic number is $\Nt\sim 10.5$, which gives
$N \simeq 5.4 \times 10^{55}$.

The variation of the chemical potential results in a whole class
of solutions depicted by the dashed line in fig. 
\ref{fig5}.
We display the mass to radius relation $M_{tot}(R)$. This function 
has a maximum at $R \simeq 25$ km, which separates two branches
of the curve.
The left branch is denoted by
a thinner dotted line in fig. (\ref{fig5}) because it corresponds to 
unstable configurations. For a given value of the total  
fermionic number $N$ there are two solutions with different 
values of $M_{tot}$ \cite{tdlee1,lynn}.
The dotted line corresponds to the branch
with larger mass. 
The solutions of figs. \ref{fig3} and \ref{fig4} are denoted by a
circle and a square, respectively, on this plot.

If the value of the scale $\sx$ changes while $\phi_1$ is kept constant,
the line $M_{tot}(R)$ retains its shape, but is rescaled by 
an overall factor. This is obvious from the form of eqs. 
(\ref{eq1d})--(\ref{rhod}), in which $\sx$ does not appear
explicitly after the rescaling of eq. (\ref{dimension}), 
as it has been incorporated in the various dimensionless
parameters. 
If a solution with given values of $\sx$, $\phi_1$ and $\mu_0$ is
known, another solution can be generated through the replacements
$\sx\to \alpha \sx$, 
$\phi \to \phi$, 
$r\to r/\alpha^2$, $\rho \to \alpha^4 \rho$,
$p \to \alpha^4 p$ and
$\phi_1 \to \phi_1$, $\mu_0 \to \alpha\, \mu_0$,
$R \to R/\alpha^2$, $M_{tot} \to M_{tot}/\alpha^2$. 
In fig. \ref{fig5} we present a class of solutions with 
$\phi_1/M$=1, $\sx/$GeV=3/2 (solid curve) that demonstrates this point. 
Each point of this curve can be obtained from a point of the 
curve with $\phi_1/M$=1, $\sx/$GeV=1 (dashed curve) by dividing the coordinates
by a factor 9/4. Lowering $\sx$ results in compact objects whose 
maximum mass can be significantly larger than the solar mass, while their
average density falls $\sim \sx^4$. 

Changing the asymptotic value 
$\phi_1$ of the field, while keeping the scale $\sx$ constant, 
modifies the shape of the 
curve $M_{tot}(R)$. In fig. \ref{fig5} we plot 
$M_{tot}(R)$ for 
$\phi_1/M$=3/2, $\sx/$GeV=1 (short-dashed curve).
Comparison with the case $\phi_1/M$=1, $\sx/$GeV=1 (dashed curve)
shows that the maximum mass is significantly reduced, while
the mass is a decreasing function of the radius for the whole 
stable branch. The form of the curve is very reminiscent of that of  
conventional neutron stars. The reason is that for large $\phi_1$ 
the fractional deviation in the interior of the
solution from the asymptotic field value is not very large. 
The fermionic mass is reduced 
but does not approach zero. This is similar to the case of nuclear matter,
for which the effective reduction of the
mass of a free fermion ($\sim 1$ GeV) is of the 
order of the nuclear binding energy ($\sim 15$ MeV).
In fig. \ref{fig5} we also depict the case with 
$\phi_1/M$=3/2, $\sx/$GeV=2/3 (dot-dashed curve).
This has the same asymptotic fermionic mass as the case
$\phi_1/M$=1, $\sx/$GeV=1 (dashed curve).

It is interesting that the upper part of the
stable branch of the cases with $\phi_1/M=1$ has the same form as
for fermion soliton stars \cite{tdlee1}, or fermion Q-stars
\cite{lynn}, or strange stars \cite{strangestars}. 
For these cases the fermionic mass approaches zero
in the interior and the binding energy is large, similarly to the
fermion stars. 
The mass is an increasing function of the radius. 
However, the lower part of the stable branch differs from that of the 
fermion stars. The mass becomes a decreasing function of the radius, 
similarly to the conventional neutron stars.

\section{Summary and conclusions}
\label{conclusions}
\setcounter{equation}{0}

In this work we considered some of 
the implications for astrophysical configurations
composed of dark matter of the possible interaction
with a scalar field that is responsible for the dark energy of the 
Universe. We assumed that at cosmological 
scales the scalar field has an almost constant expectation value. 
The time dependence associated with the evolution of the
dark energy density was neglected, as we assumed 
that the relevant time scale is 
very long. The interaction between dark matter and
dark energy was modelled by assuming that the the mass of the 
dark matter particles depends on the expectation value of the scalar field. 
We developed a formalism based on the Thomas-Fermi approximation, which
assumes that 
in a local frame at every point in space the particles 
are described by a Fermi-Dirac distribution with 
position-dependent temperature and chemical potential. 
We derived the Einstein equations and 
the equation of motion of the scalar field.

In regions of high number density of dark matter particles, the 
scalar field is shifted from its cosmological value. The shift is
in the direction that reduces the dark matter mass, so that the
total energy of the configuration is minimized. This deviation from
the asymptotic value is the manifestation of the attractive 
interaction between the dark matter particles 
mediated by the scalar field. 

We discussed the implications for the dark matter halos that comprise
the bulk of the matter outside the core of galaxies, up to distances of
100--200 kpc. 
Our approach is based on the assumption that the dark matter can be
treated as a gas of thermalized particles. It may seem unlikely that the
formalism we developed could apply to the dark matter halos, as the
particles in them interact very weakly. However, we showed that
our formalism is equivalent to the model of the isothermal sphere, which
gives a simple explanation for the approximately flat rotation curves in
the absence of the scalar interaction. The effective temperature 
in our approach
is proportional to the dispersion  $ \langle \vrm_d^2 \rangle$ 
of the velocity of the dark matter particles.

The presence of the new attractive force does not
modify the distribution of dark matter so as to destroy the 
approximately flat rotation curves. 
The main new
effect is that the velocity of a massive object orbiting 
the galaxy outside its core is not of the order of the typical velocity of
the dark matter particles, as in the conventional picture. Instead, it is
reduced by a factor $\left( 1+\kappa^2 \right)^{1/2}$, where 
$\kappa^2=4M^2\left[ (dm_0/d\phi)/{m_0}\right]^2$ quantifies the 
dependence of the dark matter mass $m_0$ on the scalar field.
If $\kappa^2$ is large, the typical velocity of the dark matter particles
can be significantly larger than the rotation velocity. 
The latter quantity is directly measurable, and its value is
used in order to deduce the velocity of dark matter particles 
for dark matter searches. 

For $\kappa^2 \gta 1$ the typical velocity of dark matter in our neighborhood
of the Milky Way exceeds the rotation velocity significantly. 
The flux of dark matter particles towards a terrestrial
detector is larger roughly by a factor $(1+\kappa^2)^{1/2}$ relative to the 
$\kappa^2=0$ case. 
As a result, the counting rates are increased
by the same factor.
Existing bounds on dark matter properties from direct searches
can be extended to include the case of non-zero $\kappa$.
The bound on the cross section for the interaction  
of dark matter with the material of the detector is strengthened by 
the factor $(1+\kappa^2)^{1/2}$. 
Corrections to this simple picture are also possible 
through the velocity dependence of various contributions to
the counting rates, such as nuclear form factors.
A separate study is required in order to determine the implications
for the various experiments.

Using the same formalism we also derived solutions that describe
denser compact objects composed of dark matter. In the examples we
discussed, the dark matter gas has zero temperature and the stability
of the configurations is provided by the degeneracy pressure.
The particle mass is determined by the scalar field through a
Yukawa interaction.
The compact objects resemble neutron stars, but 
are composed of dark matter, while their density can deviate significantly
from nuclear density. 
Similarly to the case of dark matter halos, the 
potential of the scalar field plays a minor role in the structure of
these objects. 

Their mass to radius curve has a stable branch whose
shape depends on the asymptotic value of the scalar field in Planck units
$\phi_1/M$.
For $\phi_1/M\gta 1$ the mass of the compact object 
is a decreasing function of the radius, and
the curve resembles strongly the one for
neutron stars. 
The reason is that for large $\phi_1$ 
the fractional deviation in the interior of the
solution from the asymptotic field value is not very large.
The fermionic mass is reduced 
but does not approach zero. This is similar to the case of nuclear matter.
For $\phi_1/M\lta 1$
the upper part of the
stable branch has the same form as
for fermion soliton stars \cite{tdlee1}, or fermion Q-stars
\cite{lynn}, or strange stars \cite{strangestars}. 
For these cases the fermionic mass approaches zero
in the interior and the binding energy is large.
The mass is an increasing function of the radius. 
However, the lower part of the stable branch differs from that of 
fermion stars. The mass becomes a decreasing function of the radius, 
similarly to neutron stars. 

The mass to radius curve also depends 
on the 
scale $\sx$ that determines the mass of the dark
matter particles. We assumed that the particle 
mass is generated by the scalar field through 
a Yukawa term, with a Planck suppressed Yukawa coupling $\sx/M$.
For $\phi_1/M={\cal O}(1)$ and
$\sx={\cal O}(1)$ GeV, 
the most massive astrophysical object
has $M_{tot}={\cal O}(1)\,M_\odot$ and $R={\cal O}(10)\,\,{\rm km}$.
Changing $\sx\to \alpha \sx$ leads to the rescaling   
$R \to R/\alpha^2$, $M_{tot} \to M_{tot}/\alpha^2$. 
Reducing the scale $\sx$ results in very massive, but
dilute astrophysical objects.
For $\sx={\cal O} (10)$ keV, we have 
$M_{tot}={\cal O}(10^{10})\,M_\odot$ and $R={\cal O}(10^{10})\,\,{\rm km}$.
These objects are similar to the supermassive neutrino stars hypothesized
in ref. \cite{bilic}. 
For $\sigma ={\cal O} (10^{-3})$ eV the radius approaches the horizon size. 
Conversely, increasing $\sx$ gives rise to very dense configurations of 
smaller size. For $\sx ={\cal O}(1)$ TeV and $\phi_1 ={\cal O}(1)\,M$
the mass of an unbound dark matter particle is ${\cal O}(1)$ TeV. 
The most massive compact object composed of dark matter particles
has $M_{tot}={\cal O}(10^{-6})\,M_\odot$ and $R={\cal O}(1)\,{\rm cm}$.

The configurations we described in this work correspond only to 
a subclass of the 
possible solutions for the system of a Thomas-Fermi gas interacting 
with a scalar field in the presence of gravity. 
We considered an interaction only through the fermionic mass. 
Also, the novel astrophysical objects we described
were assumed to have zero temperature. This assumption 
is probably unrealistic, as
it means that during gravitational collapse
the dark matter particles lose most of their momentum, apart from the amount 
required in order to satisfy the exclusion principle.

Modifying some of our assumptions leads to different solutions. 
For example, 
a non-zero value for the temperature of the dark matter gas
is expected to result in more dilute 
objects of larger size. 
Our aim here has been to demonstrate that the presence of an interaction
between the dark matter and the dark energy has many more observable
consequences, other than the modification of the cosmological evolution.
One interesting property of our solutions
is that the potential of the 
scalar field does not play any significant role.
The reason is that the resolution of the coincidence problem
requires the characteristic scale of this potential to be of the order
of the Hubble scale. As a result, the potential is negligible at length
scales smaller than the horizon.

Some of the ingredients of the model we considered can be fixed by establishing
the correct theory underlying the dark sector. The form of the dependence of
the dark matter mass on the scalar field, 
the value of the parameter $\sx$ that sets the scale for this mass, 
the type of the scalar potential are typical examples. On the other hand,
there are parameters that can be determined only by studying the
dynamical mechanism that leads to the formation of astrophysical
objects composed of dark matter. One such quantity is the temperature
of the fermionic gas. This could be either an effective temperature of 
a weakly-interacting gas, as in the case of galaxy halos, or the 
real temperature of a thermalized gas.

Addressing the dynamical problem is beyond the scope of this work. 
The crucial question is whether the gas of dark matter particles 
can lose momentum fast enough for dense objects to develop. 
The dark matter is expected to have only weak interactions, while its
coupling to the scalar field is suppressed by the Planck scale. 
Despite this, it has been demonstrated that massive astrophysical
objects can form under gravitational collapse \cite{bilic,diemand},
even in the absence of the attraction mediated by the scalar field. 
For light neutrinos it has been shown in ref. \cite{bilic} 
that matter can be expelled in a series of bounces during the
collapse, so that a condensed object is left behind.
In ref. \cite{diemand} the case of neutralinos with a mass of 100 GeV
has been considered. Through numerical simulations it has been
shown that compact objects as large as the
solar system and with a mass of the order of the Earth mass
start forming at redshifts $z\simeq 60$ and survive until today.  

The presence of an additional scalar attractive force increases
the likelihood of formation of compact objects.
The details of their structure (distribution on the 
radius--mass plane, density profile), the probability of their formation and
the resulting abundance
can be obtained only through a dynamical calculation that goes beyond
linearized gravity within a specific theory of the dark sector. 
An analytical approach could make use of the model of spherical 
collapse, which has already been used for the study of structure
formation for coupled dark energy and dark matter \cite{spherical}.
 
The most promising possibility for the detection of the astrophysical objects
we discussed is through gravitational microlensing. 
The sensitivity depends crucially on  
the ratio of the object radius to the
Einstein radius
\be
\frac{R_E}{{\rm km}}\simeq 0.37 \times 10^{12} \sqrt{\frac{M_{tot}}{M_\odot}}
\label{einst} \ee
for gravitational lensing. The zero-temperature solutions that we presented
have a mass to radius curve approximately given by eqs. 
(\ref{physr}), (\ref{massr}). For $\sx \gta 10^{-3}$ eV they satisfy 
$R < R_E$, so that the corresponding
astrophysical objects are expected to 
produce a detectable lensing signal. This must be contrasted with 
the much more dilute dark matter subhalos in the absence of an attractive
force mediated by a scalar field \cite{diemand}. These have size
larger than the Einstein radius and are more difficult to detect.

\vspace {0.5cm}
\noindent{\bf Acknowledgments}\\
\noindent 
We would like to thank Z. Berezhiani, P. Salucci and 
C. Wetterich for useful discussions. 
This work was supported by
the RTN contract MRTN--CT--2004--503369 of the European Union, 
the research programs 
``Pythagoras I'' (grant 70-03-7315) and 
``Pythagoras II'' (grant 70-03-7992) 
of the Greek Ministry of National Education, partially funded by the
European Union, 
and the research program ``Kapodistrias'' of the University of Athens.


\vskip 1.5cm

\begin{thebibliography}{999}



\bibitem{peeblesb}
  P.~J.~E.~Peebles,
  ``Principles of Physical Cosmology'', 
  Princeton University Press (1993).


\bibitem{peeblesrev}
  P.~J.~E.~Peebles and B.~Ratra,
  %``The cosmological constant and dark energy,''
  Rev.\ Mod.\ Phys.\  {\bf 75} (2003) 559
  [arXiv:astro-ph/0207347].
  %%CITATION = ASTRO-PH 0207347;%%

\bibitem{supernova}
  A.~G.~Riess {\it et al.}  [Supernova Search Team Collaboration],
  %``Observational Evidence from Supernovae for an Accelerating Universe and a
  %Cosmological Constant,''
  Astron.\ J.\  {\bf 116} (1998) 1009
  [arXiv:astro-ph/9805201]; \\
  %%CITATION = ASTRO-PH 9805201;%%
  S.~Perlmutter {\it et al.}  [Supernova Cosmology Project Collaboration],
  %``Measurements of Omega and Lambda from 42 High-Redshift Supernovae,''
  Astrophys.\ J.\  {\bf 517} (1999) 565
  [arXiv:astro-ph/9812133]; \\
  %%CITATION = ASTRO-PH 9812133;%%
  W.~J.~Percival {\it et al.}  [The 2dFGRS Collaboration],
  %``The 2dF Galaxy Redshift Survey: The power spectrum and the matter content
  %of the universe,''
  Mon.\ Not.\ Roy.\ Astron.\ Soc.\  {\bf 327} (2001) 1297
  [arXiv:astro-ph/0105252].
  %%CITATION = ASTRO-PH 0105252;%%


\bibitem{wetdil}
  C.~Wetterich,
  %``Cosmology And The Fate Of Dilatation Symmetry,''
  Nucl.\ Phys.\ B {\bf 302} (1988) 668.
  %%CITATION = NUPHA,B302,668;%%

\bibitem{peeblesold}
  P.~J.~E.~Peebles and B.~Ratra,
  %``Cosmology With A Time Variable Cosmological 'Constant',''
  Astrophys.\ J.\  {\bf 325} (1988) L17.
  %%CITATION = ASJOA,325,L17;%%

\bibitem{wetcosmon}
  C.~Wetterich,
  %``The Cosmon model for an asymptotically vanishing time dependent
  %cosmological 'constant',''
  Astron.\ Astrophys.\  {\bf 301} (1995) 321
  [arXiv:hep-th/9408025].
  %%CITATION = HEP-TH 9408025;%%

\bibitem{amendola}
  L.~Amendola,
  %``Coupled quintessence,''
  Phys.\ Rev.\ D {\bf 62} (2000) 043511
  [arXiv:astro-ph/9908023].
  %%CITATION = ASTRO-PH 9908023;%%

\bibitem{peeblesfar}
  G.~R.~Farrar and P.~J.~E.~Peebles,
  %``Interacting dark matter and dark energy,''
  Astrophys.\ J.\  {\bf 604} (2004) 1
  [arXiv:astro-ph/0307316].
  %%CITATION = ASTRO-PH 0307316;%%

\bibitem{mainini}
  R.~Mainini and S.~A.~Bonometto,
  %``Dark matter and dark energy from the solution of the strong CP-problem,''
  Phys.\ Rev.\ Lett.\  {\bf 93} (2004) 121301
  [arXiv:astro-ph/0406114];
  %%CITATION = ASTRO-PH 0406114;%%
\\
  R.~Mainini, L.~P.~L.~Colombo and S.~A.~Bonometto,
  %``Dark matter and dark energy from a single scalar field and CMB data,''
  arXiv:astro-ph/0503036.
  %%CITATION = ASTRO-PH 0503036;%%

\bibitem{massimo}
  M.~Pietroni,
  %``Dark energy condensation,''
  Phys.\ Rev.\ D {\bf 72} (2005) 043535
  [arXiv:astro-ph/0505615].
  %%CITATION = ASTRO-PH 0505615;%%


\bibitem{spec}
  T.~Biswas and A.~Mazumdar,
  %``Can we have a stringy origin behind Omega(Lambda)(t) proportional to
  %Omega(m)(t)?,''
  arXiv:hep-th/0408026; \\
  %%CITATION = HEP-TH 0408026;%%
  T.~Biswas, R.~Brandenberger, A.~Mazumdar and T.~Multamaki,
  %``Current acceleration from dilaton and stringy cold dark matter,''
  arXiv:hep-th/0507199.
  %%CITATION = HEP-TH 0507199;%%


\bibitem{radiative}
  S.~M.~Carroll,
  %``Quintessence and the rest of the world,''
  Phys.\ Rev.\ Lett.\  {\bf 81} (1998) 3067
  [arXiv:astro-ph/9806099];\\
  %%CITATION = ASTRO-PH 9806099;%%
  C.~F.~Kolda and D.~H.~Lyth,
  %``Quintessential difficulties,''
  Phys.\ Lett.\ B {\bf 458} (1999) 197
  [arXiv:hep-ph/9811375];\\
  %%CITATION = HEP-PH 9811375;%%
  M.~Doran and J.~Jaeckel,
  %``Loop corrections to scalar quintessence potentials,''
  Phys.\ Rev.\ D {\bf 66} (2002) 043519
  [arXiv:astro-ph/0203018];\\
  %%CITATION = ASTRO-PH 0203018;%%
  R.~Barbieri, L.~J.~Hall, S.~J.~Oliver and A.~Strumia,
  %``Dark energy and right-handed neutrinos,''
  arXiv:hep-ph/0505124.
  %%CITATION = HEP-PH 0505124;%%

\bibitem{largescale}
  B.~A.~Gradwohl and J.~A.~Frieman,
  %``Dark matter, long range forces, and large scale structure,''
  Astrophys.\ J.\  {\bf 398} (1992) 407; \\
  %%CITATION = ASJOA,398,407;%%
  L.~Amendola and D.~Tocchini-Valentini,
  %``Baryon bias and structure formation in an accelerating universe,''
  Phys.\ Rev.\ D {\bf 66} (2002) 043528
  [arXiv:astro-ph/0111535].
  %%CITATION = ASTRO-PH 0111535;%%

\bibitem{maccio}
  A.~V.~Maccio, C.~Quercellini, R.~Mainini, L.~Amendola and S.~A.~Bonometto,
  %``N-body simulations for coupled dark energy: halo mass function and density
  %profiles,''
  Phys.\ Rev.\ D {\bf 69} (2004) 123516
  [arXiv:astro-ph/0309671].
  %%CITATION = ASTRO-PH 0309671;%%

\bibitem{peeblessim}
  A.~Nusser, S.~S.~Gubser and P.~J.~E.~Peebles,
  %``Structure Formation With a Long-Range Scalar Dark Matter Interaction,''
  Phys.\ Rev.\ D {\bf 71} (2005) 083505
  [arXiv:astro-ph/0412586].
  %%CITATION = ASTRO-PH 0412586;%%

\bibitem{wetl}
  C.~Wetterich,
  %``Are galaxies cosmon lumps?,''
  Phys.\ Lett.\ B {\bf 522} (2001) 5
  [arXiv:astro-ph/0108411].
  %%CITATION = ASTRO-PH 0108411;%%

\bibitem{damour}
  T.~Damour and A.~M.~Polyakov,
  %``The String dilaton and a least coupling principle,''
  Nucl.\ Phys.\ B {\bf 423} (1994) 532
  [arXiv:hep-th/9401069].
  %%CITATION = HEP-TH 9401069;%%

\bibitem{tdlee1}
  T.~D.~Lee and Y.~Pang,
  %``Fermion Soliton Stars And Black Holes,''
  Phys.\ Rev.\ D {\bf 35} (1987) 3678.
  %%CITATION = PHRVA,D35,3678;%%

\bibitem{lynn}
  S.~Bahcall, B.~W.~Lynn and S.~B.~Selipsky,
  %``Fermion Q Stars,''
  Nucl.\ Phys.\ B {\bf 325} (1989) 606;
  %%CITATION = NUPHA,B325,606;%%
%  S.~Bahcall, B.~W.~Lynn and S.~B.~Selipsky,
  %``Are Neutron Stars Q Stars?,''
  Nucl.\ Phys.\ B {\bf 331} (1990) 67;\\
  %%CITATION = NUPHA,B331,67;%%
  B.~W.~Lynn, A.~E.~Nelson and N.~Tetradis,
  %``Strange Baryon Matter,''
  Nucl.\ Phys.\ B {\bf 345} (1990) 186.
  %%CITATION = NUPHA,B345,186;%%

\bibitem{bilic}
  N.~Bilic and R.~D.~Viollier,
  %``Gravitational phase transition of fermionic matter in a
  %general-relativistic framework,''
  Eur.\ Phys.\ J.\ C {\bf 11} (1999) 173
  [arXiv:hep-ph/9809563];
  %%CITATION = HEP-PH 9809563;%%
%  N.~Bilic and R.~D.~Viollier,
  %``General-relativistic Thomas-Fermi model,''
  Gen.\ Rel.\ Grav.\  {\bf 31} (1999) 1105
  [arXiv:gr-qc/9903034];\\
  %%CITATION = GR-QC 9903034;%%
  N.~Bilic, R.~J.~Lindebaum, G.~B.~Tupper and R.~D.~Viollier,
  %``On the formation of degenerate heavy neutrino stars,''
  Phys.\ Lett.\ B {\bf 515} (2001) 105
  [arXiv:astro-ph/0106209].
  %%CITATION = ASTRO-PH 0106209;%%

\bibitem{silk}
  G.~Bertone, D.~Hooper and J.~Silk,
  %``Particle dark matter: Evidence, candidates and constraints,''
  Phys.\ Rept.\  {\bf 405} (2005) 279
  [arXiv:hep-ph/0404175].
  %%CITATION = HEP-PH 0404175;%%

\bibitem{nfw}
  G.~Gentile, P.~Salucci, U.~Klein, D.~Vergani and P.~Kalberla,
  %``The cored distribution of dark matter in spiral galaxies,''
  Mon.\ Not.\ Roy.\ Astron.\ Soc.\  {\bf 351} (2004) 903
  [arXiv:astro-ph/0403154]; \\
  %%CITATION = ASTRO-PH 0403154;%%
  A.~Burkert,
  %``The Structure of dark matter halos in dwarf galaxies,''
  IAU Symp.\  {\bf 171} (1996) 175
  [Astrophys.\ J.\  {\bf 447} (1995) L25]
  [arXiv:astro-ph/9504041]; \\
  %%CITATION = ASTRO-PH 9504041;%%
  J.~F.~Navarro, C.~S.~Frenk and S.~D.~M.~White,
  %``The Structure of Cold Dark Matter Halos,''
  Astrophys.\ J.\  {\bf 462} (1996) 563
  [arXiv:astro-ph/9508025]; 
  %%CITATION = ASTRO-PH 9508025;%%
  %``A Universal density profile from hierarchical clustering,''
  Astrophys.\ J.\  {\bf 490} (1997) 493; \\
  %%CITATION = ASJOA,490,493;%%
  B.~Moore, F.~Governato, T.~Quinn, J.~Stadel and G.~Lake,
  %``Resolving the Structure of Cold Dark Matter Halos,''
  Astrophys.\ J.\  {\bf 499} (1998) L5
  [arXiv:astro-ph/9709051].
  %%CITATION = ASTRO-PH 9709051;%%



\bibitem{hhalos}
  N.~Tetradis,
  %``Dark Energy and Dark Matter in Galaxy Halos,''
  arXiv:hep-ph/0507288.
  %%CITATION = HEP-PH 0507288;%%

\bibitem{factors}
  J.~D.~Lewin and P.~F.~Smith,
  %``Review of mathematics, numerical factors, and corrections for dark  matter
  %experiments based on elastic nuclear recoil,''
  Astropart.\ Phys.\  {\bf 6} (1996) 87.
  %%CITATION = APHYE,6,87;%%

\bibitem{spergel}
  A.~K.~Drukier, K.~Freese and D.~N.~Spergel,
  %``Detecting Cold Dark Matter Candidates,''
  Phys.\ Rev.\ D {\bf 33} (1986) 3495.
  %%CITATION = PHRVA,D33,3495;%%

\bibitem{witten}
  M.~W.~Goodman and E.~Witten,
  %``Detectability Of Certain Dark-Matter Candidates,''
  Phys.\ Rev.\ D {\bf 31} (1985) 3059.
  %%CITATION = PHRVA,D31,3059;%%

\bibitem{strangestars}
  C.~Alcock, E.~Farhi and A.~Olinto,
  %``Strange Stars,''
  Astrophys.\ J.\  {\bf 310} (1986) 261.
  %%CITATION = ASJOA,310,261;%%


\bibitem{diemand}
  J.~Diemand, B.~Moore and J.~Stadel,
  %``Earth-mass dark-matter haloes as the first structures in the early
  %universe,''
  Nature {\bf 433} (2005) 389.
  %%CITATION = NATUA,433,389;%%


\bibitem{spherical}
  D.~F.~Mota and C.~van de Bruck,
  %``On the spherical collapse model in dark energy cosmologies,''
  Astron.\ Astrophys.\  {\bf 421} (2004) 71
  [arXiv:astro-ph/0401504];
  %%CITATION = ASTRO-PH 0401504;%%
\\
  N.~J.~Nunes and D.~F.~Mota,
  %``Structure Formation in Inhomogeneous Dark Energy Models,''
  arXiv:astro-ph/0409481;
  %%CITATION = ASTRO-PH 0409481;%%
%\cite{Manera:2005ct}
\\
  M.~Manera and D.~F.~Mota,
  %``Cluster number counts dependence on dark energy inhomogeneities and
  %coupling to dark matter,''
  arXiv:astro-ph/0504519.
  %%CITATION = ASTRO-PH 0504519;%%



\end{thebibliography}
\end{document}